\documentclass[12pt,thmsa]{article}
\usepackage{amssymb}

\usepackage{sw20aip}

\input{tcilatex}
\begin{document}

\author{Pawel O.\ Mazur\thanks{%
E-mail address: mazur@mail.psc.sc.edu} \\
Department of Physics and Astronomy,\\
University of South Carolina, Columbia, S. C. 29208, U. S. A.}
\title{$\Bbb{BLACK}$ $\Bbb{HOLE}$ $\Bbb{UNIQUENESS}$ $\Bbb{THEOREMS}$\thanks{%
This review is based on my plenary lecture at GR11 in Stockholm, July 1986.
The plenary lecture was published in \emph{Proceedings of the 11th
International Conference on General Relativity and Gravitation, }ed. M. A.
H. MacCallum, Cambridge University Press, Cambridge 1987, pp. 130-157.}}
\maketitle

\begin{abstract}
I review the \emph{black hole uniqueness theorem} and the \emph{no hair
theorems }established for \emph{physical} black hole stationary states by
the early 80'. This review presents the original and decisive work of
Carter, Robinson, Mazur and Bunting on the problem of no bifurcation and
uniqueness of \emph{physical }black holes. Its original version was written
only few years after my proof of the Kerr-Newman et al. black hole
uniqueness theorem \cite{M82a} has appeared in print. The proof of the black
hole uniqueness theorem relies heavily on the positivity properties of
nonlinear sigma models on the Riemannian noncompact symmetric spaces with
negative sectional curvature. It is hoped that the first hand description of
the original developments leading to our current understanding of the black
hole uniqueness will be found useful to all interested in the subject.
\medskip 
\end{abstract}

\section{Introduction}

It is well known that spherical gravitational collapse produces a black hole
(Oppenheimer and Snyder, 1939) \cite{OS39}. There are reasons to believe
that this is true for collapse with small deviations from spherical symmetry
(Doroshkevich, Zel'dovich and Novikov, 1966) \cite{DZN66} and it was
conjectured by Penrose (1969) \cite{Pen69} that no `naked singularities' can
occur during complete gravitational collapse (Cosmic Censorship Conjecture).
The Cosmic Censorship Conjecture is also the fundamental and not yet proved
assumption we make when studying black hole equilibrium states. The
considerable body of work on perturbations of stationary black holes brought
us the presently accepted picture of gravitational collapse (Regge and
Wheeler 1957; Visheshwara 1970; Ipser 1971; Price 1972; Press and Teukolsky
1973; Wald 1973, and others) \cite{RW57,Vish70,I71,Price72,PT73,W73}. One
would expect that a black hole formed by gravitational collapse would settle
down to a stationary equilibrium state. Properties of equilibrium states
were extensively studied in the late sixties and early seventies (Carter,
1971, 1973; Bardeen, Carter and Hawking, 1973; Hawking and Ellis, 1973;
Hawking, 1973) \cite{C71,C73,BCH73,HE73,Hawk73}. Theorems of Israel, Carter,
Hawking and Robinson obtained between 1967 and 1975 gave proof of the
remarkable result that Kerr (1963) \cite{Kerr63} back holes are the only
possible stationary vacuum black holes. The first black hole uniqueness
theorem came as a surprise when Israel (1967, 1968) \cite{Israel67,Israel68}
proved that a static, topologically spherical black hole is described by the
Schwarzschild or the Reissner-N\"{o}rdstr\"{o}m solutions. No hair and
uniqueness theorems for a stationary axisymmetric, topologically spherical
black hole were obtained by Carter (1971, 1973) \cite{C71,C73}, Robinson
(1974, 1975) \cite{RDC74,RDC75}, Mazur (1982) \cite{M82a,M82b,M84a,M84b}
(and independently by Bunting (1983) \cite{Bun83}) using the Ernst (1968) 
\cite{E68} and Geroch (1971) \cite{G71} formulation of the Einstein
equations for a stationary and axisymmetric gravitational field equations
with a nice positivity property.

\smallskip

The basic assumptions made in a proof of uniqueness theorems were justified
by Hawking (1973) \cite{Hawk73} who demonstrated that a stationary black
hole must be static or axisymmetric and the horizon has a spherical
topology. Also, Hajicek (1973) \cite{Hajicek73} has shown that the outer
boundary of the ergosphere must always intersect the event horizon. All
these results depend on the validity of the Cosmic Censorship Hypothesis.
Causality of a part of spacetime outside the event horizon is also assumed
in a proof of uniqueness theorems.

\smallskip

In this lecture I describe recent work on black hole uniqueness theorems,
reformulating the problem of uniqueness of the Kerr-Newman et al. black hole
solutions as a problem in the harmonic map theory. I would like to refer the
reader to the excellent review article by Carter (1979) \cite{C79} for a
more detailed account of basic assumptions made in a proof of uniqueness
theorems.

\section{The Einstein-Maxwell Equations for Spacetimes with One Killing
Vector}

The Ernst-Geroch formulation of the Einstein-Maxwell equations for the
stationary axisymmetric case can be carried out with respect to one of the
two existing Killing vectors. It is crucial for the proof of black hole
uniqueness theorems to have such a formulation which leads to the harmonic
maps between two Riemannian spaces, because only then may one hope to have a
global divergence identity with the required positivity property. One can
define, therefore, the Ernst potential with respect to the stationary
Killing vector $\partial /\partial t$ or with respect to the axial symmetry
Killing vector $\partial /\partial \phi .$ However, as we will see below, it
is the second choice which leads to a Riemannian metric on the image space
of the Ernst potential because the norm $X=\left( \partial /\partial \phi
,\partial /\partial \phi \right) $ is everywhere positive and vanishes only
on the symmetry axis. $X$ is positive because we assume that the black hole
spacetimes are causal in the domain of outer communications (Carter, 1979) 
\cite{C79}. What happens when we use the locally timelike Killing vector $%
\partial /\partial t?$ First of all, we notice that the norm $-V=\left(
\partial /\partial t,\partial /\partial t\right) $ of $\partial /\partial t$
changes sign on the ergosurface of rotating black holes. This means that the
signature of the metric on the image space $\mathbf{N}$ of the Ernst
potentials is changing as we cross the ergosurface of the black hole. The
metric on $\mathbf{N}$ is pseudoriemannian inside the ergosphere. For this
reason the global divergence identity is losing its nice positivity property
and besides the difficulties with deriving boundary conditions for black
holes, in this case, we cannot apply it in the proof of the black hole
uniqueness conjecture!\smallskip

It may be useful to give a short presentation of a reduction of E-M
equations from 4-dim to 3-dim using a covariant Geroch formulation for the
case with one Killing vector $\xi =\xi ^{a}\partial _{a}.$ Let $\left( 
\mathcal{M},g_{ab}\right) ,F_{ab}$ be solutions to the Einstein-Maxwell
equations

\begin{equation}
R_{ab}-\frac{1}{2}Rg_{ab}=8\pi T_{ab},  \label{one}
\end{equation}

\begin{equation}
D_{b}F^{ab}=0,  \label{two}
\end{equation}

\begin{equation}
D_{[a}F_{bc]}=0,  \label{three}
\end{equation}
where

\begin{equation}
T_{ab}=\left( 4\pi \right) ^{-1}\left( F_{a}^{c}F_{bc}-\frac{1}{4}%
g_{ab}F_{cd}F^{cd}\right) ,  \label{four}
\end{equation}
and $F_{ab}$ is the electromagnetic strength tensor.

\smallskip

Assume that $\left( g_{ab},F_{ab}\right) $are invariant with respect to a
one-parameter isometry group generated by the Killing vector field $\xi =\xi
^{a}\partial _{a}$, i. e., $\frak{L}_{\xi }g=\frak{L}_{\xi }F=0.$ Our
convention for the signature of $g_{ab}$ is $\left( -+++\right) .$ The
projection of the E-M equation on the space of orbits of $\xi $ is
straightforward (we follow Geroch formulation and notation (Geroch 1971) 
\cite{G71}). The metric on the space of orbits $M$ is $h_{ab}=g_{ab}-\lambda
^{-1}\xi _{a}\xi _{b},$ where $\lambda =\xi _{a}\xi ^{a}.$ At this point it
is convenient to notice that when $\xi $ is timelike, then $h_{ab}$ is
Riemannian; on the other hand, if $\xi $ is spacelike, as is the case with $%
\partial /\partial t$ Killing vector inside the ergosphere of a black hole,
then $h_{ab}$ is pseudoriemannian and the reduced field equations are
hyperbolic in this region! One can also define a twist vector $\widetilde{%
\omega }_{a}$ which is non-vanishing if $\xi $ is not hypersurface
orthogonal Killing vector, $\widetilde{\omega }_{a}=\epsilon _{abcd}\xi
^{b}D^{c}\xi ^{d}.$ In the vacuum case knowing $\left( M,h_{ab},\lambda ,%
\widetilde{\omega }_{a}\right) $ one can reconstruct $\left( \mathcal{M}%
,g_{ab}\right) $ completely. In the electrovacuum case one needs more
information, i. e., we have to know the `electric' and `magnetic' fields
defined as follows: $E^{a}=-F_{ab}\xi ^{b},$ $B_{a}=-^{*}F_{ab}\xi ^{b},$%
which are automatically defined on $M$, i. e., $E_{a}\xi ^{a}=B_{a}\xi
^{a}=0 $. Knowing $E_{a}$, $B_{a}$ one can reconstruct $F_{ab}:$

\begin{equation}
F_{ab}=\lambda ^{-1}\left( 2\xi _{[a}E_{b]}-\epsilon _{abcd}\xi
^{c}B^{d}\right) .  \label{five}
\end{equation}
It can be easily seen (Carter, 1972) \cite{C73} that ($\ref{two})$, $(\ref
{three})$ and the Killing equations on $\left( g,F\right) $ imply the
existence of `electric' and `magnetic' potentials $E$ and $B$ in a simply
connected region of $M,$ i. e., $E_{a}=\nabla _{a}E,$ $B_{a}=\nabla _{a}B.$
The basic equations for spacetimes with one Killling vector $\xi $ which we
apply in derivation of the reduced E-M equations are the following (we
present them here for the sake of completeness; it can be seen that using
them and the E-M equations, one arrives easily at the final Ernst form of
the field equations. In the following, we will only indicate the main steps
in the derivation of the reduced equations.)

\begin{equation}
D_{a}\xi _{b}=\frac{1}{2}\lambda ^{-1}\epsilon _{abcd}\xi ^{c}\widetilde{%
\omega }^{d}+\lambda ^{-1}\xi _{[b}\nabla _{a]}\lambda ,  \label{six}
\end{equation}

\begin{equation}
D_{a}D_{b}\xi _{c}=\xi ^{d}R_{dabc},  \label{seven}
\end{equation}

\begin{equation}
\nabla _{[a}\widetilde{\omega }_{b]}=-\epsilon _{abcd}\xi ^{c}R_{e}^{d}\xi
^{e},  \label{eight}
\end{equation}

\begin{equation}
\nabla _{a}\widetilde{\omega }^{a}=\frac{3}{2}\lambda ^{-1}\widetilde{\omega 
}^{a}\nabla _{a}\lambda ,  \label{nine}
\end{equation}

\begin{equation}
\nabla _{a}\nabla ^{a}\lambda =\frac{1}{2}\lambda ^{-1}\nabla _{a}\lambda
\nabla ^{a}\lambda -\lambda ^{-1}\widetilde{\omega }_{a}\widetilde{\omega }%
^{a}-2R_{ab}\xi ^{a}\xi ^{b},  \label{ten}
\end{equation}

\begin{equation}
^{(3)}R_{ab}=\frac{1}{2}\lambda ^{-2}\left( \widetilde{\omega }_{a}%
\widetilde{\omega }_{b}-h_{ab}\widetilde{\omega }_{c}\widetilde{\omega }%
^{c}\right) +\frac{1}{2}\lambda ^{-1}\nabla _{a}\nabla _{b}\lambda -\frac{1}{%
4}\lambda ^{-2}\nabla _{a}\lambda \nabla _{b}\lambda
+h_{a}^{c}h_{b}^{d}R_{cd},  \label{eleven}
\end{equation}
where $^{(3)}R_{ab}$ is the Ricci tensor of $h_{ab}.$ Using the field
equations (\ref{one}), (\ref{four}) and equations (\ref{seven}), (\ref{ten}%
), we arrive at

\begin{equation}
\nabla _{[a}\widetilde{\omega }_{b]}=-4E_{[a}B_{b]}.  \label{twelve}
\end{equation}
It is easy to see that we can define a curl-less `twist' vector

\begin{equation}
\omega _{a}=-\widetilde{\omega }_{a}+2\left( BE_{a}-EB_{a}\right) ,
\label{thirteen}
\end{equation}
i. e., $\nabla _{[a}\omega _{b]}=0.$ In a simply connected region one can
define, therefore, a potential $\omega :$ $\omega _{a}=\nabla _{a}\omega .$
It is easy to check that if we conformally rescale the metric $h_{ab}$ using
a scale factor depending on $\lambda $ we can cancel the second derivatives
of $\lambda $ in the expression for the Ricci tensor on the manifold of
orbits. The required conformal transformation has the form: $h_{ab}=\lambda
^{-1}\gamma _{ab}.\medskip \medskip $

Introducing the complex Ernst potentials $\epsilon $ and $\psi $ as follows

\begin{equation}
\epsilon =-\lambda -\psi \overline{\psi }+i\omega ,\text{ }\psi =E+iB,
\label{fourteen}
\end{equation}
and taking all covariant derivatives $\nabla _{a}$ with respect to $\gamma
_{ab},$ we can write equations (\ref{eleven}), (\ref{twelve}) and (\ref
{thirteen}) in a nice symmetric form of 3-dim Einstein equations coupled to
the Ernst equations:

\begin{equation}
-\lambda \nabla _{a}\nabla ^{a}\epsilon =\left( \nabla _{a}\epsilon +2%
\overline{\psi }\nabla _{a}\psi \right) \nabla ^{a}\epsilon ,
\label{fifteen}
\end{equation}

\begin{equation}
-\lambda \nabla _{a}\nabla ^{a}\psi =\left( \nabla _{a}\epsilon +2\overline{%
\psi }\nabla _{a}\psi \right) \nabla ^{a}\psi ,  \label{sixteen}
\end{equation}

\begin{equation}
R_{ab}\left( \gamma \right) =\frac{1}{2}\lambda ^{-2}\left[ \nabla
_{(a}\epsilon \nabla _{b)}\overline{\epsilon }+2\psi \nabla _{(a}\epsilon
\nabla _{b)}\overline{\psi }+2\overline{\psi }\nabla _{(a}\overline{\epsilon 
}\nabla _{b)}\psi +4\lambda \nabla _{(a}\psi \nabla _{b)}\overline{\psi }%
\right] .  \label{seventeen}
\end{equation}
Now it can be easily seen that equations (\ref{fifteen}), (\ref{sixteen})
and (\ref{seventeen}) can be derived from the action principle of the 3-dim
gravity coupled to a harmonic map model

\begin{equation}
S=\int \mid \gamma \mid ^{\frac{1}{2}}d^{3}x\left( R-L\right) ,
\label{eighteen}
\end{equation}
where $R$ is the Ricci scalar of the metric $\gamma _{ab}$ and $L$ is given
by

\begin{equation}
L=\frac{1}{2}\lambda ^{-2}\left[ \mid \nabla \epsilon +2\overline{\psi }%
\nabla \psi \mid ^{2}+4\mid \nabla \psi \mid ^{2}\right] .  \label{nineteen}
\end{equation}
$L$ defines an `energy' integral for a harmonic map between two spaces $%
\left( M,\gamma _{ab}\right) $and $\left( N,H_{AB}\right) ,$i. e., a map
between the space of orbits $M$ of $\xi $ with its conformal metric $\gamma
_{ab}$ and the space $N$ of Ernst's potentials $\Phi ^{A}=\left( \epsilon
,\psi \right) $with the metric $H_{AB}$ as defined by $L.$ At this point let
me recall the definition of the \emph{harmonic map.} I will give here a
definition using coordinate systems on two spaces $M$ and $N,$ but one can
easily see that it is a coordinate invariant statement.\bigskip

\begin{definition}
Let $\left( M,\gamma \right) $ and $\left( N,H\right) $ be two
(pseudo)Riemannian manifolds with metrics $\gamma $ and $H$ respectively. A
map $g:$ $M\rightarrow N$ which in local coordinate systems $x^{a}$ and $%
\Phi ^{A}$ on $M$ and $N,$ respectively, can be written as: $\Phi
^{A}=g^{A}\left( x^{a}\right) $ is called a \emph{harmonic map} if the
energy functional 
\begin{equation}
I\left[ g\right] =\frac{1}{2}\int_{M}\mid \gamma \mid ^{\frac{1}{2}}dx\gamma
^{ab}H_{AB}\left( \Phi \right) \nabla _{a}\Phi ^{A}\nabla _{b}\Phi ^{B},
\label{twenty}
\end{equation}
is stable under small deformations of $g,$ i. e., if the Lagrange-Euler
equations for the functional $I\left[ g\right] $are satisfied (Bochner,
1940; Fuller, 1954; Eells and Sampson, 1964) \cite{Boch40,F54,ES64}. $N$ is
called the image space of the harmonic map $g$.\medskip \medskip 
\end{definition}

The Lagrangian (\ref{nineteen}) defines the hermitian metric on the image
space $N$ of the Ernst potentials $\epsilon $ and $\psi $

\begin{equation}
ds^{2}=\frac{1}{2}\left( \func{Re}\epsilon +\psi \overline{\psi }\right)
^{-2}\left[ \mid d\epsilon +2\overline{\psi }d\psi \mid ^{2}-4\left( \func{Re%
}\epsilon +\psi \overline{\psi }\right) d\psi d\overline{\psi }\right] .
\label{twentyone}
\end{equation}
It is quite interesting to study properties of this metric. The metric (\ref
{twentyone}) \ is Riemannian only when $\lambda =-\left( \func{Re}\epsilon
+\psi \overline{\psi }\right) >0;$otherwise it is a pseudoriemannian metric.
It will be demonstrated later that the image space $N$ is in fact the
symmetric homogeneous space $N=SU\left( 2,1\right) /S\left( U\left( 2\right)
\times U\left( 1\right) \right) $ when $\lambda >0,$ and $N=SU\left(
2,1\right) /S\left( U\left( 1,1\right) \times U\left( 1\right) \right) $
when $\lambda <0$ (Mazur, 1982a,b, 1984a,b) \cite{M82a,M82b,M84a,M84b}.
Harmonic maps onto homogeneous spaces $G/H$ are known in physics as
nonlinear sigma models. This peculiar sigma model property of the E-M field
equations with one Killing vector proved to be crucial in proving black hole
uniqueness theorems for stationary and axisymmetric black holes. Because we
are interested in stationary and axisymmetric solutions we have to consider
further reduction of the E-M field equations. We assume that there is
another Killing vector $\eta =\eta ^{a}\partial _{a}$ commuting with the
previous Killing vector $\xi =\xi ^{a}\partial _{a},$ i. e., $\left[ \xi
,\eta \right] =0.$ Subsequent $3\rightarrow 2$ reduction from the space of
orbits of $\xi $ to the space of orbits of $\xi $ and $\eta $ is
straightforward. The Ernst potentials $\epsilon $ and $\psi $ defined with
respect to $\xi $ will be well defined on the space of orbits $M^{\prime }=%
\mathcal{M}/G_{2},$where $G_{2}$ is the abelian isometry group generated by $%
\xi $ and $\eta ,$ only if the action of $G_{2}$ on $\mathcal{M}$ has the
orthogonal transitivity property. It means that there exists a family of
2-surfaces orthogonal to 2-surfaces of transitivity of $G_{2}.$ The
dimensional $4\rightarrow 2$ reduction of the E-M equations from $4$ to $2$
dimensions can be carried out if the reduction does not depend on the order
of the reduction in two steps. We first consider $4\rightarrow 3$ reduction
with respect to $\xi $ and later $3\rightarrow 2$ reduction with respect to $%
\eta .$ The result of this reduction should be the same if we change the
order of subsequent steps, i. e., we first reduce with respect to $\eta $
and later with respect to $\xi .$ The consistency conditions for the $%
4\rightarrow 2$ reduction are, of course, the vanishing of Lie derivatives
of $\epsilon $ and $\psi $ with respect to the second Killing vector, i. e., 
$\mathcal{L}_{\eta }\epsilon =$ $\mathcal{L}_{\eta }\psi =0,$where $\epsilon 
$ and $\psi $ are defined with respect to $\xi .$ Analogous conditions
should be satisfied for the reversed order of the two steps of reduction;
they are equivalent to the Frobenius integrability condition for the
orthogonal transitivity of $G_{2}.$ As a result of reduction from $4$ to $2$
dimensions, we will obtain a $2$-dimensional gravity. Because any $2$%
-dimensional manifold is conformally flat and a nonlinear sigma model is
conformally invariant in two dimensions, we observe that nonlinear sigma
model equations will decouple from the two-dimensional gravity. It means
that we can solve first nonlinear sigma model equations on the $2$%
-dimensional manifold of orbits $M^{\prime }$ choosing the flat metric on $%
M^{\prime }.$ The remaining $2$-dim Einstein equations will determine the
metric $g_{\mu \nu }=e^{2\Gamma }\delta _{\mu \nu },$ $\mu =1,2$, i. e., its
conformal scale $\Gamma $ on $M^{\prime }$ (we have chosen a `conformal
gauge' on $M^{\prime }$). Once the solution to nonlinear sigma model
equations is known we can determine the metric $g_{\mu \nu }$ on $M^{\prime }
$ and we can reconstruct the four-dimensional spacetime $\left( \mathcal{M}%
,g_{ab}\right) $completely. One can see, therefore, that the problem of
uniqueness of black hole solutions of the Einstein-Maxwell equations can be
studied as a two dimensional boundary value problem for the nonlinear sigma
model equations.

\section{Black Hole Boundary Conditions}

To this end let me present the black hole boundary value problem. We refer
the reader to the more detailed derivation of these conditions discussed by
Carter (Carter, 1973, 1979) \cite{C73,C79}. We will follow Carter's
notation. Thus if $\xi =\partial /\partial \phi $ we denote $\lambda =X$ and 
$\omega =Y.$ The determinant of the matrix of scalar products of two Killing
vectors $\xi =\partial /\partial \phi ,$ $\eta =\partial /\partial t$
defines the real scalar field $\rho :-\rho ^{2}=\left( \xi ,\xi \right)
\left( \eta ,\eta \right) -\left( \xi ,\eta \right) ^{2}.$ The scalar field $%
\rho $ vanishes on the symmetry axis $\mathcal{A},$ which is a set of fixed
points of $\xi =\partial /\partial \phi $ and on the event horizon $\mathcal{%
H}.$ $\rho $ is also a harmonic function on the space of orbits $M^{\prime }$
of $\xi $ and $\eta .$ After reductions of the E-M equations for stationary
and axisymmetric fields, we obtain a set of equations on the space of orbits 
$M^{\prime }$

\begin{equation}
\nabla _{\mu }\nabla ^{\mu }\rho =0,  \label{twentytwo}
\end{equation}

\[
-X\nabla _{\mu }\left( \rho \nabla ^{\mu }\epsilon \right) =\rho \left(
\nabla _{\mu }\epsilon +2\overline{\psi }\nabla _{\mu }\psi \right) \nabla
^{\mu }\epsilon , 
\]

\begin{equation}
-X\nabla _{\mu }\left( \rho \nabla ^{\mu }\psi \right) =\rho \left( \nabla
_{\mu }\epsilon +2\overline{\psi }\nabla _{\mu }\psi \right) \nabla ^{\mu
}\psi ,  \label{twentythree}
\end{equation}
where $\nabla _{\mu }$ is the covariant derivative with respect to the
metric $g_{\mu \nu }$ on the space of orbits $M^{\prime }.\medskip $ The
black hole solutions are solutions to (\ref{twentytwo}) and (\ref
{twentythree}) for the Carter boundary conditions (Carter, 1973, 1979) \cite
{C73,C79}, which can be written explicitly once we fix a coordinate system
on $\left( M^{\prime },g_{\mu \nu }\right) .$ Once we know the topology of
the event horizon of stationary black holes is $R^{1}\times S^{2}$ (Hawking,
1972, 1973) \cite{Hawk72,Hawk73} and the topology of the domain of outer
communications is $R^{2}\times S^{2}$ we can show, using Morse theory, that
the harmonic function $\rho $ does not have \emph{critical} points on $%
\overline{M}^{\prime }.$ It means that a gradient of $\rho $ and its
harmonic conjugate function $z$ is non-zero on $\overline{M}^{\prime }$
(Carter, 1973, 1979) \cite{C73,C79}. From the asymptotic flatness condition,
we have that $\rho $ at large distances behaves as the Weyl canonical
cylindrical coordinate. The size of the event horizon fixes the overall
scale of $\rho .$ We can take $\rho $ and $z,$ its harmonic conjugate
function, as the globally well behaved coordinates on $\overline{M}^{\prime
} $

\begin{equation}
\rho ^{2}=c^{2}\left( x^{2}-1\right) (1-y^{2}),\text{ }z=cxy,
\label{twentyfour}
\end{equation}
where

\[
c^{2}=M^{2}-\left( J/M\right) ^{2}-Q^{2}. 
\]
$x=1$ defines the location of the event horizon $\mathcal{H},$ $y=$ $\pm 1$
are the two branches of the symmetry axis $\mathcal{A},$ and we reach the
spatial infinity at $x\rightarrow +\infty $ . The black hole's boundary
conditions are parametrized by three parameters: the angular momentum $J$,
the electric charge $Q$ and the parameter $c$ or the total mass of a black
hole. The black hole solution is described by the asymptotically flat
spacetime which is regular on the axis $\mathcal{A}$ and on the event
horizon $\mathcal{H}.$ These conditions can be translated into the formalism
discussed above. The asymptotic boundary conditions are:

\[
E=Qy+O\left( x^{-1}\right) , 
\]

\[
B=O\left( x^{-1}\right) , 
\]

\[
Y=2Jy(3-y^{2})+O\left( x^{-1}\right) , 
\]

\begin{equation}
X=c^{2}x^{2}(1-y^{2})+O\left( x^{-1}\right) ,  \label{twentyfive}
\end{equation}
as $x\rightarrow \infty $ . The symmetry axis boundary conditions are: $E,$ $%
B,$ $X$ and $Y,$ should be well behaved functions of $x$ and $y$ which
satisfy

\[
\partial _{x}E=O(1-y^{2}),\text{ }\partial _{y}E=O\left( 1\right) , 
\]

\[
\partial _{x}B=O(1-y^{2}),\text{ }\partial _{y}B=O\left( 1\right) , 
\]

\[
\partial _{x}Y=O(1-y^{2}),\text{ }\partial _{y}Y+2(E\partial _{y}B-B\partial
_{y}E)=O(1-y^{2}), 
\]

\begin{equation}
X=O(1-y^{2}),\text{ }X^{-1}\partial _{y}X=-2y(1-y^{2})^{2}+O(1-y^{2}),
\label{twentysix}
\end{equation}
as $y\rightarrow \pm 1.$ The event horizon boundary conditions demand only
that\ $E,$ $B,$ $X$ and $Y$ are well behaved functions of $x$and $y$ as $%
x\rightarrow 1$

\[
\partial _{x}E=O(1),\text{ }\partial _{y}E=O(1), 
\]

\[
\partial _{x}B=O(1),\text{ }\partial _{y}B=O(1), 
\]

\[
\partial _{x}Y=O(1),\text{ }\partial _{y}Y=O(1), 
\]

\begin{equation}
X=O(1),\text{ }X^{-1}=O(1),  \label{twentyseven}
\end{equation}
It was conjectured that the only stationary and axisymmetric black hole
solutions satisfying these boundary conditions are described by the
Kerr-Newman et al. (Kerr, 1963; Newman et al., 1965) \cite{Kerr63,Netal65}
family of solutions satisfying the condition of $c^{2}=M^{2}-\left(
J/M\right) ^{2}-Q^{2}>0.$ Great progress in proving correctness of this
conjecture was achieved initially in the early seventies (Carter, 1971,1973;
Robinson, 1974) \cite{C71,C73,RDC74} and then it culminated in a remarkable
black hole uniqueness theorem for vacuum black holes (Robinson, 1975) \cite
{RDC75}.\medskip

The proof of \ \emph{no hair} \emph{theorems} of Carter and Robinson and
Robinson's \emph{uniqueness theorem }for the Kerr black hole made use of
remarkable divergence identities. The common ancestor of these mysterious
identities is the Green identity for the Laplace equation, which is usually
applied to show that the Dirichlet or Neuman boundary value problem is well
posed. The \emph{no hair theorems }of Carter and Robinson are basically
statements about the absence of nontrivial bifurcations for solutions of the
harmonic map equations (\ref{twentytwo}) and (\ref{twentythree})\ with the
black hole boundary conditions (\ref{twentyfive}), (\ref{twentysix}) and (%
\ref{twentyseven}). It means that if we consider the linearized harmonic map
equations for small perturbations of a harmonic map around a black hole
solution with fixed values of $M,$ $J$ and $Q$ then there are no
perturbations satisfying linearized black hole boundary conditions. The
conclusion one could draw from these results (Carter, 1973) \cite{C73} was
that if there were other solutions, they should form a disjoint family of
solutions parametrized also by three conserved quantities: the total mass $%
M, $ the total angular momentum $J$ and the electric charge $Q$. \medskip

What remained to be done was to exclude the possibility of other than the
Kerr-Newman et al. families of black hole solutions, proving, therefore, 
\emph{the black hole uniqueness theorem}. For vacuum black holes, this goal
was achieved by D. C. Robinson in 1975 who found a nonlinear version of
Carter's divergence identity and applied it to the proof of the black hole
uniqueness theorem for the Kerr black hole. The most general form of the
black hole uniqueness theorem was proven only recently\bigskip

\begin{theorem}
\emph{Black Hole Uniqueness Theorem (Mazur, 1982a,b; Bunting, 1983) }The
only possible stationary and axisymmetric black hole solutions of the
Einstein-Maxwell equations satisfying boundary conditions (\ref{twentyfive}%
), (\ref{twentysix}) and (\ref{twentyseven}) are the Kerr-Newman et al.
solutions subject to the\smallskip constraint $M^{2}-\left( J/M\right)
^{2}-Q^{2}>0.$ Black holes are \emph{completely} characterized by three
parameters only: the total mass $M,$ the total angular momentum $J$ and the
total electric charge $Q.$ If there is a magnetic charge $P$ in Nature, then
black holes will be described completely by the four parameter family of
Kerr-Newman et al. solutions with $M^{2}-\left( J/M\right)
^{2}-Q^{2}-P^{2}>0.$
\end{theorem}

\medskip

Before proceeding with the proof of this \emph{uniqueness theorem for black
holes, }I would like first to introduce very useful techniques of the
harmonic map theory. The remarkable divergence identities of Carter and
Robinson will emerge as a natural consequence of the harmonic map property
of the Einstein-Maxwell equations for stationary and axisymmetric fields.

\section{Divergence Identities for Harmonic Maps and No Hair Theorems}

Harmonic maps enjoy a lot of nice properties; especially we can say this for
harmonic maps between two Riemannian manifolds where the image manifold has
non-positive sectional curvature (Yau, 1982) \cite{Yau82}. It is well known
that if $N$ has non-positive curvature, than every map from $M$ to $N$ is
homotopic to a harmonic map with minimal `energy'. This existence theorem
for harmonic maps between two compact Riemann manifolds was established by
Eells and Sampson (1964) \cite{ES64}. Moreover, there exists one to one and
only one harmonic map in each homotopy class of maps between two compact
Riemann manifolds\ $M$ and $N,$ when $N$ has negative curvature and if the
image of the first homotopy group $\pi _{1}\left( M\right) $ of $M$ in $\pi
_{1}\left( N\right) $ is not a cyclic group (Hartman, 1967) \cite{Hartman67}%
. Mathematically, the black hole uniqueness theorem is an example of a
result which shows that under certain boundary conditions the same
uniqueness theorem for harmonic maps can be extended to the non-compact
case. I will present here results discussed first in (Mazur, 1984a,b; Mazur
TPUJ 83 preprint) \cite{M84a,M84b} (and independently by Bunting (1983) \cite
{Bun83}). Because the \emph{no hair theorems} are concerned with small
perturbations of black hole solutions, we will be concerned here with the
question of `stability' of harmonic maps to a negatively curved Riemann
manifold $N.$ The Euler-Lagrange equations for harmonic maps can be derived
from the variational principle (\ref{twenty})

\begin{equation}
\nabla _{a}\nabla ^{a}\Phi ^{A}+\Gamma _{BC}^{A}\nabla _{a}\Phi ^{B}\nabla
^{a}\Phi ^{C}=0  \label{twentyeight}
\end{equation}
For stationary and axisymmetric E-M fields eq.$\left( \ref{twentyeight}%
\right) $is modified in the way that a covariant Laplacian in $\left( \ref
{twentyeight}\right) $ is replaced by $\rho ^{-1}\nabla \left( \rho \nabla
\right) ,$ where $\rho $\ is a non-negative harmonic function. This
modification does not change conclusions we will draw about properties of
harmonic maps, because all divergence identities presented below will be
multiplied by a non-negative function $\rho $.\medskip

In order to study the problem of bifurcations off a family of black hole
solutions we need the linearized form of $\left( \ref{twentyeight}\right) .$
Let us consider a one-parameter family of maps $f_{\tau }:M\rightarrow N,$
represented locally by functions $\Phi ^{A}\left( \tau ,x\right) $. A
tangent vector $\partial /\partial \tau $\ to a curve $f_{\tau }$ with
components, $X^{A}=\partial \Phi ^{A}/\partial \tau $ satisfies the `small
perturbations' equation (the Jacobi `geodesic deviation' equation; see e. g.
(Misner, 1978) \cite{Mis78})

\begin{equation}
D_{a}D^{a}X^{A}+R_{BCD}^{A}\nabla _{a}\Phi ^{B}\nabla ^{a}\Phi ^{C}X^{D}=0,
\label{twentynine}
\end{equation}
where

\begin{equation}
D_{a}X^{A}=\nabla _{a}X^{A}+\Gamma _{BC}^{A}\nabla _{a}\Phi ^{B}X^{C}.
\label{thirty}
\end{equation}
It is a general property of nonlinear systems that the nonexistence of
linearized perturbations for a given boundary value problem is sufficient to
exclude the possibility of a corresponding bifurcating family of exact
solutions. The following divergence identity is useful to determine the
possibility of bifurcation of solutions to $\left( \ref{twentyeight}\right) $

\begin{equation}
\nabla ^{a}\left( H_{AB}X^{A}D_{a}X^{B}\right)
=H_{AB}D_{a}X^{A}D^{a}X^{B}+R_{ABCD}X^{A}\nabla _{a}\Phi ^{B}X^{C}\nabla
^{a}\Phi ^{D}.  \label{thirtyone}
\end{equation}
The divergence term on the l.h.s. of $\left( \ref{thirtyone}\right) ,$when
integrated over $M$ gives the surface term contribution which vanishes for a
large class of boundary conditions (when $M$ has a boundary $\partial M;$
for $M$\ compact it vanishes identically). The first term on the r.h.s. of $%
\left( \ref{thirtyone}\right) $is non-negative when $M$ and $N$ are
Riemannian. The second term is also non-negative only when the sectional
curvature of $N$ is non-positive. If the surface term vanishes, as happens
for black hole solutions, then it follows that each term on the r.h.s. of $%
\left( \ref{thirtyone}\right) $vanishes separately, i. e., 
\[
D_{a}X^{A}=0 
\]
and 
\[
X^{[A}\nabla _{a}\Phi ^{B]}=0. 
\]
Then from $\left( \ref{thirtyone}\right) $ it follows that

\[
\ \Lambda _{B}^{A}X^{B}\ =0,
\]
where 
\[
\Lambda _{B}^{A}=R_{BCD}^{A}\nabla _{a}\Phi ^{C}\nabla ^{a}\Phi ^{D}.
\]
There can exist a non-zero solution to the zero eigen-value problem for the
matrix $\Lambda $\ if its determinant is zero. In general, a given solution
of $\left( \ref{twentyeight}\right) $for which we are seeking bifurcations
has the property that $\Lambda $\ is a non-degenerate matrix. One can also
see that the number of bifurcations is equal to $\dim (N)-rank(\Lambda ).$We
conclude that, in general, there are no bifurcations for harmonic maps when
the sectional curvature of $N$ is non-positive (unless we have a degenerate
case). The Carter and Robinson \emph{no hair theorems} are simply very
special cases of the result I have described briefly above. The positive
definite divergence identity $\left( \ref{thirtyone}\right) $when applied to
electrovacuum black holes, i. e., to solutions of $(\ref{twentytwo})$ and (%
\ref{twentythree}) satisfying black hole boundary conditions (\ref
{twentyfive}), (\ref{twentysix}) and (\ref{twentyseven}) gives as a result $%
\emph{the}$ $\emph{black}$ $\emph{hole}$ $\emph{no}$ $\emph{hair}$ $\emph{%
theorem}$ (Carter, 1971, 1973; Robinson, 1974; Mazur, 1984a,b) \cite
{C71,C73,RDC74,M84a,M84b)}. This is so because the image space of the Ernst
potentials $\epsilon $ and $\psi $\ is a Riemannian space with negative
sectional curvature (Mazur, 1982a,b, 1984a,b) \cite{M82a,M82b,M84a,M84b}. It
is quite interesting that the complicated divergence identities of Carter
and Robinson are just special cases of the general divergence identity $%
\left( \ref{thirtyone}\right) $for perturbations of harmonic maps to
negatively curved Riemann manifold $N.$ It is obvious that using a
particular Ernst's coordinate system on $\left( N,H_{AB}\right) ,$ as given
by (\ref{twentyone}), one can obtain a rather unenlightening form of the
identity $\left( \ref{thirtyone}\right) $. It is the advantage of the
geometrical approach of harmonic map theory which could bring us a better
understanding of global black hole solutions. We have, therefore, a nice
theorem\medskip \medskip 

\begin{theorem}
\emph{No Hair Theorem }(Carter, 1971, Robinson, 1974) There do not exist
regular, small perturbations of the; stationary and axisymmetric black hole
solutions of the E-M equations preserving boundary conditions (\ref
{twentyfive}), (\ref{twentysix}) and (\ref{twentyseven}) for fixed values of
the mass $M,$ the angular momentum $J$ and the electric charge $Q$ of a
black hole. The only deformations of black holes that do exist are those
obtained by a change of $M,$ $J,$ and $Q.$
\end{theorem}

\medskip

\smallskip It is also easy to see that starting with the divergence identity
for small perturbations of harmonic maps one can obtain the \emph{global }%
divergence identity with the required positivity property when $N$ has a
non-positive sectional curvature. Consider a curve of harmonic maps $f_{\tau
}$ , $\tau \in \left[ 0,1\right] $\ . Then $f_{0}$ and $f_{1}$ are
homotopically equivalent. For each \ the r.h.s. of $\left( \ref{thirtyone}%
\right) $is non-negative. If we integrate both sides of the divergence
identity $\left( \ref{thirtyone}\right) $along a geodesic curve $f_{\tau }$
the right hand side (r.h.s.) will be non-negative. The left hand side
(l.h.s.) will have a form

\begin{equation}
\int\limits_{0}^{1}d\tau \nabla _{a}\nabla ^{a}\left(
H_{AB}X^{A}X^{B}\right) =\nabla _{a}\nabla ^{a}\int\limits_{0}^{1}d\tau
\left( \partial /\partial \tau ,\partial /\partial \tau \right) \geqslant 0.
\label{thirtytwo}
\end{equation}
Taking advantage of the fact that there always exists a unique geodesic
joining two points $f_{0}$ and $f_{1}$ on a simply connected Riemannian
manifold $N$ with a negative sectional curvature one can evaluate the
integral $\left( \ref{thirtytwo}\right) $. Because $f_{\tau }$ is a
geodesic, it means that the norm $\left( \partial /\partial \tau ,\partial
/\partial \tau \right) $ is a constant along $f_{\tau },$ i. e., $\left(
\partial /\partial \tau ,\partial /\partial \tau \right) =c^{2}=cons\tan t.$
One can easily see that the geodesic distance between $f_{0}$ and $f_{1}$ is:

\[
S\left( f_{0},f_{1}\right) =\int\limits_{0}^{1}d\tau \sqrt{\left( \partial
/\partial \tau ,\partial /\partial \tau \right) }=c\int\limits_{0}^{1}d\tau
=c. 
\]
It means that the $\tau $\ integral in $\left( \ref{thirtytwo}\right) $is
equal to $S^{2}(f_{0},$ $f_{1}).$ In this way we have arrived at the Bunting
divergence inequality (Bunting, 1983) \cite{Bun83}

\begin{equation}
\nabla ^{a}\nabla _{a}S^{2}\left( f_{0},f_{1}\right) \geqslant 0.
\label{thirtythree}
\end{equation}
A similar identity where $S^{2}$ was replaced by an arbitrary function of $%
S^{2}$ with positive first and second derivatives was proposed also by the
present author (Mazur, 1984a,b; Mazur TPJU-22/83 preprint) \cite{M84a,M84b}.
One can apply the divergence identity $\left( \ref{thirtythree}\right) $to
the space $N$ of Ernst's potentials. $N$ is Riemannian and it has negative
sectional curvature. The `distance' function $S\left( f_{0},f_{1}\right) $
for two solutions of the harmonic map equations $\left( \ref{twentyeight}%
\right) $(and in the special case of Ernst equations (\ref{twentythree}))
satisfying the same boundary conditions is non-negative and vanishes only on
the boundary. For the case of Ernst's equations for the stationary and
axisymmetric fields the divergence identity is modified by the presence of
the harmonic, positive function $\rho $\ :

\[
\nabla _{\mu }\left( \rho \nabla ^{\mu }S^{2}\left( f_{0},f_{1}\right)
\right) \geqslant 0.
\]
When we integrate this identity over the manifold of orbits $M^{\prime }$ we
obtain surface terms which vanish. The contribution from the symmetry axis $%
\mathcal{A}$ and the event horizon $\mathcal{H}$ vanishes in virtue of the
boundary conditions $\left( \ref{twentysix}\right) $and $\left( \ref
{twentyseven}\right) $and because $\rho $ is vanishing there. The asymptotic
boundary conditions $\left( \ref{twentyfive}\right) $imply vanishing of a
surface integral at infinity where $\rho \rightarrow \infty $. The r.h.s. of
the identity $\left( \ref{thirtythree}\right) $vanishes only when\ $S\left(
f_{0},f_{1}\right) $ is constant everywhere. But $S\left( f_{0},f_{1}\right) 
$ vanishes on the boundary, so it must vanish everywhere. If we take $f_{0}$
to be the Kerr-Newman et al. black hole solution and for $f_{1}$ we take
another possible black hole solution, then vanishing of $S\left(
f_{0},f_{1}\right) $ implies $f_{0}\equiv $ $f_{1},$ i. e., \emph{the black
hole uniqueness theorem }(Mazur, 1982a,b; Bunting, 1983) \cite
{M82a,M82b,Bun83}. We have seen that the \emph{no hair theorems} and the 
\emph{black hole uniqueness theorem} can be proved applying a generalization
of the Green divergence identity to the case of harmonic maps to Riemannian
manifolds with non-positive sectional curvature. As it was mentioned above,
one can easily generalize the inequality $\left( \ref{thirtythree}\right) $%
by taking an even function of $S$ with positive first and second derivatives
with respect to $S$. Consider a function of $\sigma =S^{2},$ $h\left( \sigma
\right) ,$ and evaluate its Laplacian

\[
\nabla ^{2}h\left( \sigma \right) =h^{\prime }\left( \sigma \right) \nabla
^{2}\sigma +h^{\prime \prime }\left( \sigma \right) \left( \nabla \sigma
\right) ^{2}.
\]
Then 
\[
\nabla ^{2}h(\sigma )\geqslant 0,
\]
if $h^{\prime }\left( \sigma \right) >0,$ $h^{\prime \prime }\left( \sigma
\right) >0$ and $\left( \ref{thirtythree}\right) $is satisfied, i. e., if $N$
has non-positive sectional curvature (Mazur, 1982b, 1983, 1984a,b)\cite
{M82b,M84a,M84b}. \medskip 

The method the present author used in his proof of \emph{the black hole
uniqueness theorem} is based on a divergence identity for the Ernst
equations (\ref{twentythree}), whose derivation was based on the observation
that the image manifold of Ernst's potentials is a homogeneous, symmetric K%
\"{a}hler manifold $N=SU(2,1)/S(U(2)\times U(1)).$ This method seems to be
much simpler in application because for the electrovacuum black holes the
Riemannian space of Ernst's potentials $\epsilon $\ and $\psi $\ is a
symmetric space with the $SU(2,1)$ isometry group. Because of this large
symmetry, one can use group theoretical methods to construct an $SU(2,1)$
invariant two-point function $h\left( f_{0},f_{1}\right) $ which is a
certain function of the geodesic distance between two harmonic maps $f_{0}$
and\ $f_{1}.$ The algebraic explicit construction of the global divergence
identity offers a simple understanding of the otherwise mysterious Robinson
identity (Mazur, 1982a,b; Mazur, 1984a,b) \cite{M82a,M82b,M84a,M84b},
because the identity produced this way reproduces in the vacuum case the
Robinson identity (Robinson, 1975) \cite{RDC75}. The construction I am going
to describe here has a much broader range of applications than solely the
black hole uniqueness theorems. It can be applied to any nonlinear elliptic
sigma model on symmetric Riemannian spaces with non-compact isometry groups,
or, which is equivalent, with negative sectional curvature (Mazur, 1984a,b;
Mazur and Richter, 1985; Breitenlohner and Maison, 1986) \cite
{M84a,M84b,MR85,BM86}.

\section{A Global Identity for Nonlinear Sigma Models and It Applications:
Black Hole Uniqueness Theorem}

Before discussing the most general case of sigma models on symmetric
Riemannian and hyperbolic spaces and the divergence identity associated with
them, let me first demonstrate explicitly that the metric (\ref{twentyone})
related to the electrovacuum Ernst equations is the left invariant metric on
the homogeneous, symmetric and Kahler space $N=SU(2,1)/S(U(2)\times U(1)).$
The detailed construction is described in (Mazur, 1982a,b, 1983) \cite
{M82a,M82b,M83}. There are many ways to see this. The metric (\ref{twentyone}%
) is an example of a Hermitian metric on an (almost) complex manifold $N$
with complex coordinates $z^{\alpha }$ (in the case of electrovacuum Ernst
equations, \ $\alpha =1,2,$ $z^{1}=$ $\epsilon ,$ $z^{2}=\psi $ ): 
\[
ds^{2}=k_{\alpha \overline{\beta }}dz^{\alpha }dz^{\overline{\beta }}. 
\]
A complex Hermitian manifold $N$ is a K\"{a}hler manifold if it admits a
closed non-degenerate $\left( 1,1\right) $ form

\[
\omega =-\frac{i}{2}\ k_{\alpha \overline{\beta }}dz^{\alpha }\wedge dz^{%
\overline{\beta }}. 
\]
One can easily see that the metric (\ref{twentyone}) is K\"{a}hler (Mazur
1983) \cite{M83}, i. e.,

\[
\partial _{\gamma }k_{\alpha \overline{\beta }}=\partial _{\overline{\alpha }%
}k_{\gamma \overline{\beta }},\text{ }\partial _{\overline{\gamma }%
}k_{\alpha \overline{\beta }}=\partial _{\overline{\beta }}k_{\alpha 
\overline{\gamma }}, 
\]
where $\partial _{\alpha }=\partial /\partial z^{\alpha }$ and $\partial _{%
\overline{\alpha }}=\partial /\partial \overline{z}^{a}$. Moreover, the
metric (\ref{twentyone}) has $8$ holomorphic Killing vectors which generate
the Lie algebra of the pseudounitary group $SU(2,1).$ The group $G=SU(2,1)$
acts on $N$ nonlinearly by holomorphic (homographic) maps. One can see that
by introducing new Ernst coordinates $w^{\alpha }$ on $N:$

\[
z^{1}=\frac{w^{1}-1}{w^{1}+1},\text{ }z^{2}=\frac{w^{2}}{w^{1}+1}\text{ .} 
\]
and showing that the following holomorphic transformations of $w^{\alpha }$
are invariant transformations of the metric (\ref{twentyone})

\[
w^{1^{\prime }}=\frac{u_{1}^{1}w^{1}+u_{2}^{1}w^{2}+u_{3}^{1}}{%
u_{1}^{3}w^{1}+u_{2}^{3}w^{2}+u_{3}^{3}}\text{ ,} 
\]

\[
w^{2^{\prime }}=\frac{u_{1}^{2}w^{1}+u_{2}^{2}w^{2}+u_{3}^{2}}{%
u_{1}^{3}w^{1}+u_{2}^{3}w^{2}+u_{3}^{3}}\text{ ,} 
\]
where the matrix $u_{\beta }^{\alpha },$ $\alpha ,\beta =1,2,3$ satisfies
the pseudo-unitarity condition 
\begin{equation}
u^{+}u=\eta \text{ , }\eta =diag(1,1,-1)  \label{thirtyfour}
\end{equation}
The group $G=SU(2,1)$ acts on $N$ simply transitively. This means that $N$
is a homogeneous Kahler manifold. Every homogeneous manifold of $G$ is
diffeomorphic to the left coset space $G/H$ for some subgroup $H$ of $G.$ In
our case, it can be easily seen that $H=S(U(2)\times U(1)).$ Fix a point $%
p_{0}\in N,$ such that $w^{\alpha }(p_{0})=0.$ Every point $p\in N$ can be
reached from $p_{0}$ by a map $u:$ $p=up_{0},$ $u\in G.$ Two maps $%
u,u^{\prime }\in G$ represent the same point if $u\sim u^{\prime }=uh,$ $%
h\in H.$ $H$ is the isotropy subgroup of $p_{0}$ which is isomorphic to $%
S(U(2)\times U(1)).$ The equivalence relation $\sim $ defines the coset
space $G/H.$ The homogeneous space $N=SU(2,1)/S(U(2)\times U(1))$ is also a
symmetric space. The last property helps to construct a unique left $G$
invariant metric on $N$ in terms of a coset representative $g.$ \medskip

We would like to reformulate the Ernst equations (\ref{twentythree})
associated with the metric \smallskip (\ref{twentyone}) in terms of group
theoretical objects, like a coset representative $g(uH)=g(u),$ making their $%
SU(2,1)$ covariance explicit. This will help us to construct the generalized
Robinson identity in the obvious way. Exploiting the $SU(2,1)$ covariance of
the field equations (\ref{twentythree}) one can obtain an $SU(2,1)$
invariant divergence identity for two solutions of (\ref{twentythree}). The
metric\ (\ref{twentyone}) is a left $SU(2,1)-$invariant metric on a
symmetric space $N=SU(2,1)/S(U(2)\times U(1)).$ This leads to a nonlinear
sigma model form of the Ernst equations (\ref{twentythree}), once we go from
the Ernst parametrization of $N$ to a coset space representation of $N.$ We
give here a short derivation of a global divergence identity for nonlinear
sigma models on symmetric spaces and apply it to the electrovacuum Ernst
equations.\medskip

\smallskip We define the symmetric space as a triple $\left( G,H,\mu \right) 
$ where $G$ is a connected Lie group, and $H$ is a closed subgroup of $G$
defined by an involutive automorphism $\mu $\ of $G$ such that $(G_{\mu
})_{0}\subset H\subset G_{\mu }$ with $G_{\mu }$ and $(G_{\mu })_{0}$ being
the set of fixed points of $\mu $ and its identity component, respectively.
An involutive automorphism $\mu $\ defines also a smooth mapping $g$ of a
coset space $G/H$ into $G:$%
\[
g(u)=u\mu (u)^{-1},\text{ }g(uH)=g(u). 
\]
The $G-$valued field $g$ satisfies a constraint: 
\[
g\mu (g)=I. 
\]
In terms of $g$ one can naturally introduce the left $G-$invariant metric on 
$G/H:$%
\[
dS^{2}=\frac{1}{2}Tr(dgg^{-1})^{2}. 
\]
If $G$ is a non-compact group and $H$ is its maximally compact subgroup,
then the metric $\frac{1}{2}Tr(dgg^{-1})^{2}$ is Riemannian. A harmonic map
from a Riemannian space $M$ to a Riemannian coset space $N=G/H$ is called a 
\emph{nonlinear sigma model} with the Lagrangian

\begin{equation}
L=\frac{1}{2}Tr\left( J_{a}J^{a}\right) ,\text{ }J_{a}=\nabla _{a}gg^{-1},
\label{thirtyfive}
\end{equation}
and field equations

\begin{equation}
\nabla _{a}J^{a}=0.  \label{thirtysix}
\end{equation}
The field equation $\left( \ref{thirtysix}\right) $is covariant under the
left $G$ translation on $G/H.$ A global divergence identity for the
nonlinear sigma model $\left( \ref{thirtyfive}\right) $can be obtained
exactly in the same way as the Green identity for the Laplace equation. To
this end, consider two fields $g_{0}$ and $g_{1},$ not necessarily solutions
of $\left( \ref{thirtysix}\right) $. Define a field

\[
\Phi =g_{0}g_{1}^{-1}, 
\]
which transforms under the rigid left $G-$translations on $N,$

\[
u_{0}\rightarrow uu_{0},u_{1}\rightarrow uu_{1}, 
\]
$u\in G,$ in a simple way: \ 

\[
\Phi \ \rightarrow u\Phi u^{-1}. 
\]
One can easily see that the trace of \ \ is invariant under the rigid $G-$%
translations of two points on $N.$ It means that $Tr\Phi $ is a function of
geodesic distance between $g_{0}$ and $g_{1}:$

\[
Tr\Phi =h(S),\text{ }S=S(g_{0},g_{1})\smallskip .
\]
Evaluating the covariant Laplacian of $Tr\Phi $ and using $\left( \ref
{thirtyfive}\right) $we arrive at the generalized Green identity for
nonlinear sigma models (Mazur, 1982a,b, 1984a,b) \cite{M82a,M82b,M84a,M84b}

\begin{equation}
\nabla _{a}\nabla ^{a}h=Tr\left[ \Phi \nabla _{a}\left(
J^{(0)a}-J^{(1)a}\right) \right] +Tr\left[ \Phi \left(
J_{a}^{(0)}J^{(0)a}+J_{a}^{(1)}J^{(1)a}-2J_{a}^{(0)}J^{(1)a}\right) \right]
\label{thirtyseven}
\end{equation}
where $J_{a}^{(i)}=\nabla _{a}g_{i}g_{i}^{-1},$ $i=0,1.\bigskip $

\begin{theorem}
(Mazur, 1982b,1984a) A global identity $\left( \ref{thirtyseven}\right) $for
Riemannian nonlinear sigma models has a positivedefinite right hand side if $%
g_{0}$ and $g_{1}$ are solutions to $\left( \ref{thirtysix}\right) $and the
following conditions on $N$ are satisfied:
\end{theorem}

\begin{enumerate}
\item  $N=G/H$ is a Riemannian symmetric space with non-compact isometry
group $G$.

\item  $N$ has a non-positive sectional curvature.
\end{enumerate}

\medskip

We can see now a direct correspondence of this result to the basic
properties of harmonic mappings to a Rie\smallskip mannian space $N$ with
non-positive sectional curvature discussed before. For the electrovacuum
Ernst equations, the identity $\left( \ref{thirtyseven}\right) $is a
generalization of the Robinson identity and also it coincides with
Robinson's identity for the vacuum case. This form of the divergence
identity $\left( \ref{thirtyseven}\right) $proves to be very useful in
evaluating surface terms for black hole boundary conditions because the
function $h=Tr\Phi $\ turns out to be only a rational function of Ernst
potentials. The symmetric space $N=SU(2,1)/S(U(2)\times U(1))$ is singled
out by the existence of an inner involutive automorphism of $G=SU(2,1):$

\[
u\rightarrow \mu (u)=\eta u\eta ^{-1},\text{ }\eta =(1,1,-1), 
\]
because $\mu (H)=H,$ $H=S(U(2)\times U(1)).$ A coset representative

\[
g=u\mu (u)^{-1}, 
\]
which is an element of $G=SU(2,1),$ i. e.,

\[
g^{+}\eta g=\eta , 
\]
satisfies also a constraint

\[
g\mu (g)=g\eta g\eta ^{-1}=I. 
\]
This implies that $g$ is a hermitian matrix $g^{+}=g.$ Writing

\[
g=\eta +2P, 
\]
one can see that $P$ satisfies the following constraints:

\[
P^{+}=P,\text{ }(P\eta )^{2}=-P\eta . 
\]
We will recover the Ernst parametrization of $SU(2,1)/S(U(2)\times U(1))$
and satisfy constraints on $P$ if we write

\[
P^{\alpha \beta }=v^{\alpha }\overline{v}^{\beta }, 
\]
where

\[
v^{\alpha }\eta _{\alpha \beta }\overline{v}^{\beta }=-1,\text{ }\alpha
,\beta =1,2,3, 
\]
and define

\[
w^{1}=\frac{v^{1}}{v^{3}},\text{ }w^{2}=\frac{v^{2}}{v^{3}}. 
\]
$w^{1}$ and $w^{2}$ can be written in terms of Ernst's potentials $\epsilon $
and $\psi $\ as follows:

\[
w^{1}=\frac{1+\epsilon }{1-\epsilon }\text{ , }w^{2}=\frac{2\psi }{%
1+\epsilon }\text{ .} 
\]
Using the Ernst parametrization of the coset space element $g$ which was
discussed above\smallskip one can calculate a function $h(S)=Tr\Phi $ in
terms of the Ernst potentials $X_{i},$ $Y_{i},$ $E_{i}$ and $B_{i},$ $i=0,1$

\[
h(S)=Tr\Phi =3+X_{0}^{-1}X_{1}^{-1}\left\{ \left( X_{0}-X_{1}\right)
^{2}+2\left( X_{0}+X_{1}\right) \left[ \left( E_{0}-E_{1}\right) ^{2}+\left(
B_{0}-B_{1}\right) ^{2}\right] +\right. 
\]

\begin{equation}
\left. +\left[ \left( E_{0}-E_{1}\right) ^{2}+\left( B_{0}-B_{1}\right)
^{2}\right] ^{2}+\left[ Y_{0}-Y_{1}+2\left( E_{1}B_{0}-E_{0}B_{1}\right)
\right] ^{2}\right\}   \label{thirtyeight}
\end{equation}

The black hole uniqueness theorem for electrically charged black holes can
be obtained by a straightforward application of the divergence identity $%
\left( \ref{thirtyseven}\right) $. When we integrate the l.h.s. of $\left( 
\ref{thirtyseven}\right) $we obtain a surface term which can be shown, after
using $\left( \ref{thirtyeight}\right) $, to be vanishing for the black hole
boundary conditions (\ref{twentyfive}), (\ref{twentysix}) and (\ref
{twentyseven}). Because the r.h.s. of $\left( \ref{thirtyseven}\right) $is
positive definite, it means that $h=const.$ This is consistent with the
boundary conditions and $\left( \ref{thirtyeight}\right) $only when $%
X_{0}=X_{1},$ $Y_{0}=Y_{1},$ $E_{0}=E_{1}$ and $B_{0}=B_{1}.$ It means that
the Kerr-Newman et al. family of solutions with $M^{2}-\left( J/M\right)
^{2}-Q^{2}>0$ characterizes completely the stationary equilibrium black hole
states in the Einstein-Maxwell theory. Concluding this talk, I would like to
point out that it was only possible to prove black hole uniqueness theorems
for the stationary black holes which are also axisymmetric. It is fortunate
that the Einstein-Maxwell equations reduce to harmonic mapping equations
which are moreover conformally invariant in the case of stationary and
axisymmetric black holes. One may hope that the reasonable extension of
arguments presented here may lead to the solution of the uniqueness problem
for stationary black holes, without assumption of axial symmetry. \bigskip 

{\Large Acknowledgments}{\large \bigskip }

This work was supported in part by the national Science Foundation through
Grant No. PHY 83-18350 with Syracuse University and a travel grant, also. I
would like to thank the General Relativity Society for giving me the
opportunity to present this material. 

The address and affiliation(s) at the time this review was written were the
following: Department of \ Physics, Syracuse University, Syracuse, N.Y.
13244-1130, U. S.A. and Department of Theoretical Physics, Jagellonian
University, Krak\'{o}w, Poland.

\end{document}